\documentclass[12pt]{article}
\usepackage{amsmath,amsthm,amscd,amsfonts,amssymb}
\usepackage{latexsym}
\usepackage[usenames]{color}
\newcommand{\CC}{\mathbb C}

\newcommand{\PP}{\mathbb P}
\newcommand{\RR}{\mathbb R}

\newcommand{\ZZ}{\mathbb Z}
\newcommand{\EE}{\mathbb E}

\newcommand{\ETG}{\widetilde{E}_{0}(\Gamma)}
\newcommand{\ETc}{\widetilde{E}_{0}(\Gamma^{c})}
\newcommand{\ET}{\widetilde{E}_{0}}
\newtheorem{thm}{Theorem}[section]
\newtheorem{lemma}[thm]{Lemma}
\newtheorem{cor}[thm]{Corollary}
\newtheorem{prop}[thm]{Proposition}
\newtheorem{assum}[thm]{Assumption}

\theoremstyle{definition}
\newtheorem{rem}[thm]{Remark}
\newtheorem{defin}[thm]{Definition}

\def\half{\frac{1}{2}}

\newcommand\pf[1][]{{\noindent {\bf Proof}#1\,\bf: }}
\newcommand{\supp}{{\rm supp\,}}
\newcommand{\tr}{{\rm tr\,}}
\newcommand{\myover}[2]{\genfrac{}{}{0pt}{}{#1}{#2}}

\begin{document}
\author{
Werner Kirsch\\ Institut f\"ur Mathematik und Informatik\\ FernUniversit\"at in Hagen \\
58097 Hagen, Germany \\ email: werner.kirsch@fernuni-hagen.de\\
M Krishna \\ Ashoka Univeristy, Plot 2, Rajiv Gandhi Education City\\
Rai, Haryana 131029 India\\ email: krishna.maddaly@ashoka.edu.in  }
\title{Spectral Statistics for an Anderson Model with sporadic potentials}
\maketitle
\begin{abstract}
In this paper we consider Anderson model with a large number of sites with zero
interaction.  For such models we study the spectral statistics in the region of complete localization near the bottom of the spectrum.  We show that Poisson statistics holds for such energies, by proving the Minami estimate.
\end{abstract}

\section{Introduction}
We consider random Schr\"{o}dinger operators on $\ell^2(\ZZ^d)$ of the form
\begin{align}\label{eq:operator1}
   H_{\omega}~&=~H_{0}+\lambda\,V_{\omega}\\
\intertext{for $\lambda>0$ and with}
H_{0}u(n)~&=~(-\Delta u)(n) = -\sum_{i : |i-n|=1} \left( u(i) - u(n)\right)\label{eq:operator2}
\end{align}

Let $\tilde{V}_{\omega}(n), n\in\ZZ^{d}$ be independent and identically distributed random variables
with common distribution $\mu$. We restrict the potential $\tilde{V}_{\omega}(n)$ to a proper subset
$\Gamma$ of $\ZZ^d$ setting:
\begin{align}\label{eq:potential}
   V_{\omega}(n)~:=~\left\{
                      \begin{array}{ll}
                        \tilde{V}_{\omega}(n), & \hbox{for $n\in\Gamma$;} \\
                        0, & \hbox{otherwise.}
                      \end{array}
                    \right.
\end{align}

So, compared to the (normal) Anderson model (with potential $\tilde{V}_{\omega} $) the sites in $\Gamma^{c}:=\ZZ^d\setminus\Gamma$
are `missing' for the potential $V_{\omega}$.

In this paper, the set $\Gamma$ is periodic with respect to a sublattice $\mathbb{L}$ of $\ZZ^{d}$. For the ease of exposition we
take $\mathbb{L}= M\ZZ^{d}$, but the proofs work for more general lattices. We assume that the common distribution
$\mu$ of the $V_{\omega}(n);\,n\in\Gamma$ has a bounded density of compact support $\supp \mu$ with $\inf \supp \mu\leq 0$.
See Assumption \ref{assum} amd \ref{assumG} for the precise assumptions.

We prove that for strong disorder ($\lambda$ large) and small energies $E<E_{1}$ for some $E_{1}>0$
the level statics around $E$ is given by a Poisson process whenever the density of states $n(E)$
is positive.

Our proof follows the main ideas of Minami theory \cite{NM}. In a first step we
prove an exponential fractional moment bound for the resolvent of $H_{\omega}$. Such
bounds go back to the paper \cite{AM} (see \cite{AW} for a comprehensive treatment of the
Aizenman-Molchanov theory). This allows us to show not only Anderson localization but also that the eigenvalue counting process
is infinitely divisible in the limit. Then we prove a Minami estimate, i.\,e. we show that
this process has no double points. In a last step we have to prove
that the intensity of the limiting process is positive and, as a matter of fact, is given by the density of states $n(E)$.

To our knowledge, the first treatment of operators as in \eqref{eq:operator1}\,--\,\eqref{eq:potential} is the PhD-thesis of J\"{o}rg Obermeit \cite{JO}
which is based on ideas from \cite{WKwegner}. In \cite{JO} Obermeit proves spectral localization for small energies via multiscale analysis.

In \cite{CR} Constanza Rojas-Molina treats operators as above but with a rather general set of missing sites, in fact,  $\Gamma$ is
merely supposed to be a Delone set. One of the main technical result in \cite{CR} is a Wegner estimate for such potentials, which in turn uses methods
from \cite{CHK}.

Independently, Elgart and Klein \cite{EK} prove a similar result, they even allow a (deterministic) background potential.

The proof in Obermeit's paper \cite{JO} works if one can make sure that the energy $E$ under consideration
is outside the spectrum of a certain reference operator.
Such a reference operators may be given by $H_{0,\Gamma^{c}}:=\chi_{\Gamma^{c}}\,H_{0}\,\chi_{\Gamma^{c}}$  which is obviously non negative. This observation can be used if there is some negative spectrum of $H_{\omega}$ which is the case if $\inf \supp \mu<0$.
It is less obvious but true that $H_{0,\Gamma^{c}}\geq E_{1}>0$, so that the proof even works for small positive energies and $\inf\supp\,\mu=0$.
The estimate $H_{0,\Gamma^{c}}\geq E_{1}>0$ is proved by Elgart and Klein in \cite{EK} in great generality. In the appendix of
this paper we give an alternative proof of such a result, but with more restrictive assumptions.

In the recent work \cite{ES} Elgart and Sodin prove localization for operators as in \eqref{eq:operator1}\,--\,\eqref{eq:potential} with periodic set $\Gamma $. In this paper
the authors use the method of fractional moments (see \cite{AM} or \cite{AW}). We also use a fractional moments estimate to prove the Poisson statistics.
Under our special  assumptions the fractional moments estimate has a quite simple proof which we give below, although we could use the result from \cite{ES}.

\section{Results}
We consider Schr\"{o}dinger operators of the form \eqref{eq:operator1}--\eqref{eq:potential} with independent random variables
$V_{\omega}(n),\;n\in\Gamma$ with a common distribution $\mu$.

For the distribution $\mu$ of the random variables
we suppose:
\begin{assum}\ \label{assum}
   \begin{enumerate}
      \item The distribution $\mu$ has a bounded density and the support $\supp \mu$ of $\mu$ is compact.
      \item We have $v_{min}:=\inf\,\supp \mu~=~0$.
         \end{enumerate}
\end{assum}
A straightforward modification of our proof works also for the case $v_{min}<0$. To simplify notation we
restrict ourselves to the case $v_{min}<0$.
The set $\Gamma$ satisfies:
\begin{assum}\ \label{assumG}
   \begin{enumerate}
      \item $\Gamma~=~\Gamma_{0}\,+\,M\ZZ^{d}$\,.
      \item $\Gamma_{0}\subset \Lambda_{1}:=\{i\in\ZZ_{d}\mid -\frac{M}{2}<i_{\nu}\leq \frac{M}{2},\,\nu=1,\ldots,d\}$
      \item $\emptyset\neq \Gamma_{0}\neq \Lambda_{1}$
   \end{enumerate}
\end{assum}

We may take $\Omega = \RR^\Gamma$ as the underlying probability space with probability measure
$\PP = \prod_{i\in\Gamma} \mu$. Correspondingly,  there is an action $T$ of $\Gamma$ on
$\Omega$ by shifts and with respect to this action $\PP$ is invariant
and ergodic.   If we take the unitaries $U_{i}$ implemented by
translation on $\ell^2(\ZZ^d)$, then we have a covariance relation for
the operators $H_\omega$, namely,
\begin{equation}\label{covariance}
H_{T_i\omega} = U_i H_\omega U_i^*.
\end{equation}
It follows that the spectrum $\sigma(H_{\omega})$ is almost surely constant and the same is true for its
various measure theoretic parts (continuous spectrum, pure point spectrum etc.), see e.\,g. \cite{PF} or \cite{KiMa}.

It is well known that $\sigma(H_{0})=[0,4d]$. It follows from the above assumptions that
\begin{align}
&E_{0}~:=~\inf \sigma(H_{\omega})~=~0\qquad&&\text{if $v_{min}=\inf\supp \mu=0$}\\
\text{and}\quad&E_0~<~0 &&\text{if } v_{min}<0\,.
\end{align}

If $v_{min}<0$ then
$E_{0}$ depends on the parameter $\lambda$ from \eqref{eq:operator1},
$E_{0}$ tends to $-\infty$ as $\lambda$ goes to $\infty$ (see \cite{WK} for details).

For a subset $\Lambda$ of $\ZZ^d$ we consider
the operators
\begin{eqnarray}\label{local}
H_{\omega,\Lambda}  =  \chi_{\Lambda} H_\omega \chi_{\Lambda},
\end{eqnarray}
where $\chi_{\Lambda} $ denotes multiplication with the characteristic function of the set $\Lambda $. Analogously
we define $H_{0,\Lambda}$. Note, that we consider $H_{\omega,\Lambda}$ as an operator on $\ell^{2}(\Lambda)$.

We denote by $\Lambda_{L}(n)$ the cube of side length $LM$ centered at the origin,
more precisely
\begin{align}
   \Lambda_{L}(n)~:=~\{i\in\ZZ^d\mid -\frac{M}{2}L<(i_{\nu}-n_{\nu})\leq \frac{M}{2}L, \text{ for } \nu=1,\ldots d \}
\end{align}
and set for short $\Lambda_{L}:=\Lambda_{L}(0) $. Note that $\Lambda_{1} $ is `the' unit cell of the lattice
$M\ZZ^{d} $ and each $\Lambda_{L}(n)$ is a disjoint union of cells $\Lambda_{1}(i)$ with $i\in M\ZZ^{d}$.
The cube $\Lambda_{L}(i)$ contains exactly $|\Lambda_{L}(i)|:=(ML)^d$ points.

We denote the projection-valued spectral
measures of self-adjoint operators $B$ by $E_{B}(\cdot)$.

The measures $\nu_{L}$ with
\begin{align}\label{def:dos1}
   \nu_{L}(I)~:=~\frac{1}{|\Lambda_{L}|}\,\tr\big(E_{H_{\omega,\Lambda_{L}}}(I)\big)
\end{align}
converge weakly as $L\to\infty$ to the \emph{density of states measure} $\nu$ of $H_{\omega}$ (see e.\,g.\cite{WK}).

We remark that
$\tr A$ is always taken in the space appropriate for the operator $A$, so in \eqref{def:dos1} the trace is taken in $\ell^{2}(\Lambda_{L})\cong \CC^{|\Lambda_{L}|}$.

The density of states measure $\nu$ for an interval $I$ can also be defined by
\begin{align}\label{def:dos2}
   \nu(I)~=~\EE\Big(\frac{1}{|\Lambda_{1}|}\,\tr\big(\chi_{\Lambda_{1}}\,E_{H_{\omega}}(I)\big)\Big)
\end{align}
It is a standard result that \eqref{def:dos2} is indeed the limit of \eqref{def:dos1} (see \cite{WK}).

The following result is essential to treat not only negative energies:

\begin{prop}\label{prop:positivity}
   Under Assumption \ref{assumG} with $\Gamma^{c}=\ZZ^{d}\setminus\Gamma$
   \begin{align}
      \ETc~&:=~\inf \sigma(H_{0,\Gamma^{c}})~>~0\qquad\text{and}\\ \ETG~&:=~\inf \sigma(H_{0,\Gamma})~>~0
   \end{align}
\end{prop}
This result is proved in \cite{EK} under weaker assumptions. For the reader's convenience we give
a different proof (under our assumptions) in section \ref{sec:Appendix}.

We set
\begin{align}\label{eq:E0}
   \ET~:=\min\{\ETG,\ETc\}
\end{align}

\begin{cor}
   If $\Lambda$ is a nonempty subset of either $\Gamma$ or $\Gamma^{c}$ then
   \begin{align}
     \inf \sigma(H_{0,\Lambda})~\geq~ \ET~>~0
   \end{align}
\end{cor}
In section \ref{sec:Wegner} we prove

\begin{prop}[Wegner Estimate]\label{prop:Wegner}
   Under the assumptions \ref{assum} and \ref{assumG}
   and for given $E_{1}<\ETc$ there is a constant $C$ such that for intervals $I=[a,b]\subset [-\infty,E_{1}]$
\begin{align}\label{eq:Wegner1}
\EE \left( \tr\left(E_{H_{\omega,\Lambda}}(I) \right) \right)~ \leq~ C \,|\Lambda|\, |I|.
\end{align}

Moreover, for  energies $E<\ETc$
the density of states measure $\nu $ of the operator $H_{\omega}$ has
a bounded density $n(E)$, i.\,e. for all $E'<\ET$:
\begin{align}\label{eq:Wegner2}
   \nu\big((-\infty,E')\big)~=~\int_{-\infty}^{E'}n(E)\;
   dE
\end{align}
\end{prop}
By the Lebesgue differentiation theorem (see e.\,g. \cite{WR}, Theorem 7.7) for (Lebesgue-) almost all $E<\ET$ we have
\begin{align}
   n(E)~=~\lim_{\varepsilon\to 0}\frac{\nu\big((E-\varepsilon,E+\varepsilon)\big)}{\varepsilon}\,,
\end{align}
in particular $n(E)>0$ for $\nu$-almost all $E\in[E_{0},\ET]$.

We set
\begin{align}
   \mathcal{D}_{\nu}~=~\{E<\ET\mid \lim_{\varepsilon\to 0}\frac{\nu\big((E-\varepsilon,E+\varepsilon)\big)}{2\varepsilon} \text{ exists and is positive}\}
\end{align}
For $E\in\mathcal{D}_{\nu} $ we set $n(E)=\lim_{\varepsilon\to 0}\frac{\nu\big((E-\varepsilon,E+\varepsilon)\big)}{2\varepsilon}$. Recall that by \eqref{eq:Wegner2} $n$ is
defined only almost everywhere.

To formulate our main result we define
\begin{align}
  \xi_{L,E}^\omega(I) & = Tr\left(E_{H_{\omega, \Lambda_{L}}}(E + |{\Lambda_{L}}|^{-1}\,I)\right)\,.
\end{align}
The point process $\xi_{L,E}^\omega$ gives the local (rescaled) level statistics near energy $E$.

We prove:
\begin{thm}\label{poisson}
Suppose that the operators $H_\omega$ as in \eqref{eq:operator1}--\eqref{eq:potential} satisfy Assumptions \ref{assum} and \ref{assumG}.
Then for any $E_{1}<\ET$ there is a $\lambda$  (as in \eqref{eq:operator1}) such that:

For all $E<E_{1}$ with $E\in\mathcal{D}_{\nu}$
the limits
\begin{align}
\Xi^\omega_E(\cdot ) = \lim_{L \rightarrow \infty} \xi_{L, E}^\omega (\cdot)
\end{align}
exist in the sense of weak convergence and give a Poisson Process with intensity
measure $\lambda_E([a,b]) = n(E)(b-a) $.
\end{thm}

The proof of Theorem \ref{poisson} relies on three main ingredients which we state below and prove in the rest of this note.
Once these ingredients are given one may follow Minami's original proof \cite{NM}, see also \cite{AW}, Chapter 17.

The first ingredient is a fractional moment bound, namely:

\begin{thm}\label{fracmom}
Consider operators $H_\omega$ as in equation (\ref{eq:operator1})--\eqref{eq:potential} satisfying Assumption \ref{assum}
and \ref{assumG}.
Then for each $E_{1}<\ET$ there exist $s\in (0,\frac{1}{2})$ and $\lambda_{0}$ such that for all $\lambda\geq\lambda_{0}$ and all $E\leq E_1$ there is a $\delta>0$ with
\begin{align}
\displaystyle{ \sup_{\epsilon >0} ~ \sup_{x \in \ZZ^d} \sum_{y \in \ZZ^d} }
\EE \left(|(H_\omega - E -i\epsilon)^{-1}(x,y)|^{s}\right) e^{s\delta |x-y|} < \infty.
\end{align}
\end{thm}

Fractional moment bounds were first proved in the celebrated paper \cite{AM} by M.~Aizenman and S.~Molchanov. They imply spectral and dynamical
localization in the corresponding part of the spectrum. We refer to the book \cite{AW} by Aizenman and Warzel for a comprehensive treatment of
this area.

The above Theorem implies localization for small energies and high disorder for our model. Moreover, following N. Minami \cite{NM},
Theorem \ref{poisson} allows us to prove that the processes $\Xi$ is an infinitely devisable process by approximating the $\xi_{L}$
by sums of processes $\eta_{\ell,E, p}^\omega(I) = \tr\left(E_{H_{\omega,\ell,p}}(E + IL^{-d})\right)$ which are based on cubes
$\Lambda_{\ell}$ on a small scale $\ell$ of the order $L^{a}$ with $a<1$.

The second ingredient is a kind of extended Wegner-type estimate first proved by Minami \cite{NM}.

\begin{thm}[Minami estimate]\label{thm:Minami}
   Under the assumptions \ref{assum} and \ref{assumG} and for any given $E_{1}<\ET$ there is a
   constant $C$ such that  for intervals $I=[a,b]\subset [-\infty,E_{1}]$
\begin{align}
\EE \Big( \tr\left(E_{H_{\omega,\Lambda}}(I)\Big(\tr\left(E_{H_{\omega,\Lambda}}(I)-1\right) \Big)\right) \Big)\leq C |\Lambda|^{2} |I|^{2}\,.
\end{align}
\end{thm}
This estimate enables us to exclude double points for the limit process $\Xi$.

The final ingredient identifies the intensity measure of $\Xi$.
\begin{prop}\label{intensity}
For each $E\in\mathcal{D}_{\nu}$ and any interval $I$
\begin{align}
\lim_{L \rightarrow \infty} \EE \left(  \tr\left(\chi_{\Lambda_L} E_{H_\omega}\big(E\,+\,\frac{1}{|\Lambda_{L}|}\,I\big) \right)  \right)~ =~ n(E)\,|I|\,.
\end{align}
\end{prop}

We prove Proposition \ref{prop:Wegner}, Proposition \ref{intensity} and Theorem \ref{thm:Minami} in section \ref{sec:Wegner}. Section \ref{sec:fracmom} is devoted to
the proof of Theorem \ref{fracmom} and the appendix contains a proof of Proposition \ref{prop:positivity}.

\section{Wegner and Minami Estimates}\label{sec:Wegner}

We start this section with a lemma that allows us to reduce estimates over a cube $\Lambda$ to estimates over the `active' sites in $\Lambda$,
i.\,e. on the set $\Lambda\cap\Gamma $.

\begin{lemma}\label{uniquecont} Suppose $E_{1}<\ETc$ and
consider an eigenvalue $E\leq E_{1}$ of $H_{\omega, \Lambda}$ with $\psi$ the
corresponding eigenfunction.   Then we have for some $C$ (allowed to depend on $E_{1}$ but not on $\Lambda$)
\begin{align}
\|\psi\|~=~\|\chi_\Lambda\psi\| ~ \leq~ C\, \|\chi_{\Lambda \cap \Gamma}\,\psi\|\,.
\end{align}
\end{lemma}

\pf We have
$$
\big(H_{0, \Lambda}-E\big) \psi = -\lambda V_\omega \psi,
$$
so, since $H_{0,\Lambda}\geq \ETc>E$ we may write
$$
\psi = - (H_{0,\Lambda} - E)^{-1} \lambda V_\omega \psi.
$$
This combined with the fact that $\lambda V_\omega$ is bounded and supported
on $\Gamma$ gives the bound
\begin{align*}
   \|\psi\|_2 ~&\leq~ \|(H_{0,\Lambda} - E)^{-1}\| \|\lambda V_\omega \psi\|_2 \\
   &\leq~\|(H_{0,\Lambda} - E_{1})^{-1}\| \|\lambda V_\omega \psi\|_2 \\
   &\leq~ C\; \|\chi_{\Gamma \cap \Lambda}\psi\|.
\end{align*}
\qed

Lemma \ref{uniquecont} allows us
to bound the trace of spectral projections by the
trace over a restriction to $\Gamma$.

\begin{prop}\label{restric}
Let $I \subset (-\infty, ~ E_{1}], ~~ E_{1} < \ETc$, be an interval.
Then
$$
\tr\left( E_{H_{\omega, \Lambda}}(I)\right) \leq C\, \tr\left( \chi_{\Gamma \cap \Lambda} E_{H_{\omega, \Lambda}}(I)\right).
$$
\end{prop}
\pf We expand the trace and use Lemma \ref{uniquecont} in the second inequality,
where we write the normalized eigenfunction corresponding to an eigenvalue $\lambda$
by $\psi_\lambda$:
\begin{align}
\tr\left( E_{H_{\omega, \Lambda}}(I)\right)~&\leq \sum_{\lambda \in I} \tr\left( E_{H_{\omega, \Lambda}}(\{\lambda\})\right) = \sum_{\lambda \in I} \|\chi_\Lambda \psi_{\lambda}\|^2 \\
&
\le~  \sum_{\lambda \in I} C\; \|\chi_{\Lambda \cap \Gamma} \psi_{\lambda}\|^2 \\
&=~ C\;\tr\left(\chi_{\Lambda \cap \Gamma} E_{H_{\omega, \Lambda}}(I)\right).
\end{align}
\qed

For a given $j\in\Gamma$ we define $(\omega^{\bot}_{j},\tau)\in\RR^{\Gamma}$ by
\begin{align}
   (\omega^{\bot}_{j},\tau)(n)=\left\{
                               \begin{array}{ll}
                                 \omega_{n}, & \hbox{for $n\not=j$;} \\
                                 \tau, & \hbox{for $n=j$.}
                               \end{array}
                             \right.
\end{align}
We also denote by $\EE_{\omega^{\bot}}$ expectation over all random variable $V_{\omega}(n)$ except $V_{\omega}(j)$, so that
due to independence
\begin{align}
   \EE\Big(F(\omega)\Big)~=~\int \EE_{\omega^{\bot}}\Big(F\big((\omega^{\bot}_{j},\tau)\big)\Big)\;d\mu(\tau)
\end{align}

\begin{lemma}\label{spectralav}
Let $I \subset (-\infty, ~ E_{1}], ~~ E_{1} < \ETc$, be an interval.
Then for $j \in \Gamma \cap \Lambda$,
\begin{align}\label{eq:spectralav}
\int ~ \langle \delta_j,  E_{H_{(\omega_{j}^{\bot},\tau), \Lambda}}(I)\, \delta_j\rangle~ d\mu(\tau) ~\leq~  C'\;|I|.
\end{align}
\end{lemma}
\begin{rem}
   In \eqref{eq:spectralav} we may replace $\mu$ by any probability measure on $\RR$ with a bounded density.
\end{rem}
\pf This lemma follows by a standard spectral averaging result (see \cite{DK}, {Theorem 3.1.4}) of rank one perturbations, based on the assumptions on $\mu$. \qed

\pf[   (Proposition \ref{prop:Wegner})]
We use Lemmas \ref{restric} and \ref{spectralav} to obtain the following
series of inequalities giving \eqref{eq:Wegner1}.
\begin{align}
&\EE\left(\tr\left( E_{H_{\omega, \Lambda}}(I)\right)\right)~ \leq~ C\;\EE\left( \tr\left( \chi_{\Gamma \cap \Lambda} E_{H_{\omega, \Lambda}}(I)\right)\right) \\
 \le~& C ~ \sum_{j \in \Gamma\cap\Lambda} \EE_{(\omega_{j}^{\bot},\tau)}\left( \int d\,\mu(\tau) \langle \delta_j, ~ E_{H_{(\omega_{j}^{\bot},\tau), \Lambda}}(I) \delta_j\rangle \right) \\
 \leq~ &C'' ~~ |\Gamma \cap \Lambda | ~  |I|.
\end{align}
\eqref{eq:Wegner2} follows by an application of the Radon-Nikodym Theorem.
\qed

We turn to the proof of Minami's estimate (Theorem \ref{thm:Minami}).

\pf[ (Minami estimate)]  We first use Lemma \ref{restric} to get the bound
\begin{align*}
& \EE\Big( \tr\big(E_{H_{\omega,\Lambda}}(I)\big)\left(\tr\big(E_{H_{\omega,\Lambda}}(I)\big) - 1\right) \Big)  \\
 \leq~& C ~ \sum_{j \in \Lambda \cap \Gamma}\EE\Big( \langle \delta_j, ~ E_{H_{\omega, \Lambda}}(I) \delta_j\rangle ~ \left(\tr\big(E_{H_{\omega,\Lambda}}(I)\big) - 1\right)\Big) \\
 \leq~& C ~  \sum_{j \in \Lambda \cap \Gamma}\EE\left( \langle \delta_j, ~ E_{H_{\omega, \Lambda}}(I) \delta_j\rangle ~
\tr\big(E_{H_{(\omega_j^\perp,\tau), \Lambda}}(I)\big)\right)
\end{align*}
For the last inequality we used the fact that changing one parameter $V_{\omega}(j)$ is a rank one perturbation.
Due to eigenvalue interlacing (see e.\,g. \cite{WK} Lemma 5.25) the trace of $E_{H_{\omega, \Lambda}}(I)$ is changed
at most by one.
Since $\tau$ is a free parameter independent of anything else in the above
inequalities we integrate the inequalities over $\tau$  with
respect to $\mu$ to get
\begin{align*}
& \EE\left( \tr\big(E_{H_{\omega,\Lambda}}(I)\big)\left(\tr\big(E_{H_{\omega,\Lambda}}(I)\big) - 1\right) \right)  \\
\leq~& C ~\sum_{j \in \Lambda \cap \Gamma}\EE_{\omega_{j}^{\perp}}\left(\int d\,\mu(v) \langle \delta_j, ~
E_{H_{(\omega_{j}^{\perp},v), \Lambda}}(I) \delta_j\rangle ~
\int d\,\mu_{0}(\tau)~ \tr \big(E_{H_{(\omega_j^\perp,\tau), \Lambda}}(I)\big)\right) \\
\leq~&C''~ |I|~ \sum_{j\in\Gamma\cap\Lambda} \EE_{\omega_{j}^{\perp}}\Big(\int d\,\mu_{0}(\tau)~ \tr \big(E_{H_{(\omega_j^\perp,\tau), \Lambda}}(I)\big)\Big)\\
\leq~&C''~ |I|~ |\Lambda|~ \EE\Big( \tr \big(E_{H_{(\omega_j^\perp,\tau), \Lambda}}(I)\big) \Big)\\
\leq~& C'''~ |\Gamma \cap \Lambda|^2 ~ |I|^2,
\end{align*}
where we get first the bound for the $\tau$ integral using the spectral
averaging bound and then  the remaining integral again using Wegner estimate.
 \qed
{ }\bigskip{ }
Now, we prove Proposition \ref{intensity}.
{ }\bigskip{ }

\pf[ (Proposition \ref{intensity})]
\begin{align*}
\EE \left(  \tr\left(\chi_{\Lambda_L} E_{H_\omega - E}(\tfrac{1}{|\Lambda_{L}|}\,I) \right)  \right)
~&=~
\EE \displaystyle{\sum_{n \in \Lambda_L \cap M\ZZ^d}} \left(  \tr\left(\chi_{\Lambda_1(n)} E_{H_\omega - E}(\tfrac{1}{|\Lambda_{L}|}\,I) \right)  \right) \\
&=~
 \displaystyle{\sum_{n \in \Lambda_L \cap M\ZZ^d}}\EE \left(  \tr\left(\chi_{\Lambda_1(n)} E_{H_\omega - E}(\tfrac{1}{|\Lambda_{L}|}\,I) \right)  \right) \\
&=~
L^d~ \EE \left(  \tr\left(\chi_{\Lambda_1} E_{H_\omega - E}(\tfrac{1}{|\Lambda_{L}|}\,I) \right)  \right)  \\
&=~
L^d ~\nu(E+\tfrac{1}{|\Lambda_{L}|}\,I)\,,
\end{align*}
where we used the invariance of the expectation under translations by points
in $M\ZZ^d$, coming from the covariance equation (\ref{covariance}) and
the independence of the meausure $\PP$ under such translations of points
in $\Omega$ by $M\ZZ^d$.  Taking the limits of above we obtain the limit for $E\in\mathcal{D}_{\nu}$
$$
\lim_{L\rightarrow \infty} \EE \left( \tr\left(\chi_{\Lambda_L(0)} E_{H_\omega - E}(IL^{-d}) \right)  \right) = n(E)\;|I|\,.
$$
\qed

\section{Localization}\label{sec:fracmom}

We now turn to the proof of Theorem \ref{fracmom}, that is the proof of exponential decay of the fractional moments
of the resolvent kernels of $H_\omega$.

We start with a decoupling result.
\begin{prop}\label{prop:deco}
Suppose $\Lambda\subset\Lambda'\subset\ZZ^{d}$, $x,y\in\Lambda$ and $0<s<\half$.
Let $V:\Lambda\to\RR, V':\Lambda'\to\RR$ with $V(x)=V'(x)=0$ and set
\begin{align*}
   &H_{\Lambda}(v)~:=~H_{0,\Lambda}\,+\,V\,+\lambda\, v\; \delta_{x};\qquad H_{\Lambda'}(v)~:=~H_{0,\Lambda'}\,+\,V'\,+\lambda\, v\; \delta_{x}\,,\\
  \end{align*}
  then
\begin{align}
   &\int \Big(|(H_{\Lambda}(v)-z)^{-1}(x,y)|^{s}\,|(H_{\Lambda'}(v)-z)^{-1}(x',y')|^{s}\Big)\; d\mu(v)\notag\\
   \leq~&C\,\int |(H_{\Lambda}(v)-z)^{-1}(x,y)|^{s}\; d\mu(v)\;\int |(H_{\Lambda'}(v)-z)^{-1}(x',y')|^{s}\; d\mu(v)
\end{align}
for a constant $C$ which depends only on $s$ and the measure $\mu$.
\end{prop}

\pf The Proposition is a slight extension of Corollary 8.11 of \cite{AW}, the proof given there extends to
our case.
\qed

\medskip
\noindent Proposition \ref{prop:deco} can be iterated to give the following Corollary.
\begin{cor}\label{cor:deco}
   Under the assumptions of Proposition \ref{prop:deco}, if $y\not=x$ and $V(x)=V'(x)=V(y)=V'(y)=0$ then with
   \begin{align*}
    &H_{\Lambda}(v,u)~:=~H_{0,\Lambda}\,+\,V\,+\lambda\, v\; \delta_{x}\,+\lambda\,u\;\delta_{y}\\
    &H_{\Lambda'}(v,u)~:=~H_{0,\Lambda'}\,+\,V'\,+\lambda\, v\; \delta_{x}\,+\lambda\,u\;\delta_{y}\,,
    \end{align*}
   we have
   \begin{align*}
      \int\!\! \int \Big(&|(H_{\Lambda}(v,u)-z)^{-1}(x,y)|^{s}\,|(H_{\Lambda'}(v,u)-z)^{-1}(x',y')|^{s}\Big)\; d\mu(v)\,d\mu(u)\notag\\
   \leq\; C'&\int\!\!\! \int |(H_{\Lambda}(v,u)-z)^{-1}(x,y)|^{s}\; d\mu(v)\,d\mu(u)\;\\
   &\int\!\!\! \int|(H_{\Lambda'}-z)(v,u)^{-1}(x',y')|^{s}\; d\mu(v)\,d\mu(u)\notag
   \end{align*}
\end{cor}
We need another result from \cite{AW} (Corollary 8.4):
\begin{prop}\label{prop:est}
   If $H_{\Lambda}H=H_{\Lambda,0}+V$ and $V(x)=\lambda v$, $V(y)=\lambda u$ then there is a constant $C$
   depending only on $s$ and on the measure $\mu $ such that
   \begin{enumerate}
      \item \ \vspace{-7mm}\begin{align}
              \int |(H_{\Lambda}-z)^{-1}(x,x)|^{s}\,d\mu(v)~\leq~\frac{C}{\lambda^{s}}
            \end{align}
      \item\ \vspace{-5mm} \begin{align}
             \int \int |(H_{\Lambda}-z)^{-1}(x,y)|^{s}\,d\mu(v)\,d\mu(u)~\leq~\frac{C}{\lambda^{s}}
            \end{align}
   \end{enumerate}
\end{prop}

For the proof of Theorem \ref{fracmom} we define the following quantities:
\begin{align}
   A(z)\;&:=\;\sup_{x\in\Gamma}\;\sum_{y\in\ZZ^{d}}\EE\left(\big|\left(H_{\omega}-z\right)^{-1}(x,y)\big|^{s}\right)e^{s\delta|x-y|}\\
   B(z)\;&:=\;\sup_{x\in\Gamma^{c}}\;\sum_{y\in\ZZ^{d}}\EE\left(\big|\left(H_{\omega}-z\right)^{-1}(x,y)\big|^{s}\right)e^{s\delta|x-y|}\\
   A'(z)\;&:=\;\sup_{x\in\Gamma}\;\sum_{y\in\ZZ^{d}}\EE\left(\big|\left(H_{\omega,\Gamma}-z\right)^{-1}(x,y)\big|^{s}\right)e^{s\delta|x-y|}\\
   B'(z)\;&:=\;\sup_{x\in\Gamma^{c}}\;\sum_{y\in\ZZ^{d}}\EE\left(\big|\left(H_{\omega,\Gamma^{c}}-z\right)^{-1}(x,y)\big|^{s}\right)
   e^{s\delta|x-y|}\\
   &=\;\sup_{x\in\Gamma^{c}}\;\sum_{y\in\ZZ^{d}}\big|\left(H_{0,\Gamma^{c}}-z\right)^{-1}(x,y)\big|^{s}
   e^{s\delta|x-y|}\notag
\end{align}

We have to prove that for $E\leq E_{1}<\ET$ and $\lambda $ large enough
\begin{align}
   \sup_{\varepsilon>0}\,A(E+i\varepsilon)~<\infty \quad \text{and}\quad \sup_{\varepsilon>0}\,B(E+i\varepsilon)~<\infty
\end{align}

Since both
\begin{align}
H_{\omega,\Gamma}~&\geq ~H_{0,\Gamma}~\geq~\ET\\
\text{and}\quad H_{\omega,\Gamma^{c}}~&= ~H_{0,\Gamma^{c}}~\geq~\ET
\end{align}
the Combes-Thomas estimate (see e.\,g. \cite{AW} or \cite{WK}) shows that the matrix elements of the respective
resolvents decay exponentially uniformly in $\varepsilon $, thus
\begin{align}
   \sup_{\varepsilon>0}\,A'(E+i\varepsilon)~<\infty \quad \text{and}\quad \sup_{\varepsilon>0}\,B'(E+i\varepsilon)~<\infty
\end{align}
provided $\delta>0$ is small enough.

In the following formulae we use the abbreviations $G(x,y):=(H_{\omega}-z)^{-1}(x,y)$, $G_{0}(x,y):=(H_{0}-z)^{-1}(x,y)$,
$G_{\Gamma}(x,y):=(H_{\omega,\Gamma}-z)^{-1}(x,y)$, $G_{0,\Gamma^{c}}(x,y):=(H_{0,\Gamma^{c}}-z)^{-1}(x,y)$ and so on.

Now we estimate $B(z) $ using the geometric resolvent equation
\begin{align}\label{eq:estB}
   &B(z)~
   \leq~\sup_{x\in\Gamma^{c}}\sum_{y\in\ZZ^{d}}
   |G_{0,\Gamma^{c}}(x,y)|^{s}\,e^{s\delta|x-y|}\notag\\
   +\,&\sup_{x\in\Gamma^{c}}\sum_{y\in\ZZ^{d}}\sum_{\myover{u\in\Gamma^{c},w\in\Gamma}{|u-w|=1}}
   |G_{0,\Gamma^{c}}(x,u)|^{s}\,e^{s\delta|x-u|}\,\EE\big(|(G(w,y)|^{s}\big)\,e^{s\delta|w-y|}\,e^{s\delta}\notag\\
   \leq~& B'(z)\;+\;2\,B'(z)\, A(z)
\end{align}
where we assumed, without loss of generality, that $e^{s\delta}\leq 2$.

In a similar way we estimate
\begin{align}\label{eq:A'}
   A(z)~
   \leq~A'(z)
   +\,\sup_{x\in\Gamma}\sum_{y\in\ZZ^{d}}\sum_{\myover{u\in\Gamma^{c},w\in\Gamma}{|u-w|=1}}
   \EE\big(|G_{\Gamma}(x,w)|^{s}\,|(G(u,y)|^{s}\big)\,e^{s\delta|x-y|}
\end{align}

Using Proposition \ref{prop:deco}, Corollary \ref{cor:deco} and Proposition \ref{prop:est} we estimate:
\begin{align}
   \EE\big(|G_{\Gamma}(x,w)|^{s}\,|(G(u,y)|^{s}\big)~\leq~\frac{C  }{\lambda^{s}}\; \EE\big(|(G(u,y)|^{s}\big)\,.
\end{align}
On the other hand, due to the Combes-Thomas estimate
\begin{align}
   \EE\big(|G_{\Gamma}(x,w)|^{s}\,|(G(u,y)|^{s}\big)~\leq~C_{0}\,e^{-s\gamma|x-w|}\; \EE\big(|(G(u,y)|^{s}\big)\,.
\end{align}
Consequently, we may estimate
\begin{align}
   \sup_{x\in\Gamma} \sum_{w\in \ZZ^{d}} \EE\big(|G_{\Gamma}(x,w)|^{s}\,|(G(u,y)|^{s}\,e^{s\delta|x-y|}\big)~\leq \kappa_{\lambda} \; \EE\big(|(G(u,y)|^{s}\,e^{s\delta|u-y|}\big)
\end{align}
for a constant $\kappa_{\lambda}>0 $ which can be made arbitrarily small by choosing $\lambda $ big.

It follows that
\begin{align}\label{eq:estA}
   A(z)~\leq~A'(z)\,+\,\kappa_{\lambda}'\,B(z)\,.
\end{align}
Together with \eqref{eq:estB} this implies
\begin{align}\label{eq:estB2}
   B(z) ~\leq~B^{*}(z)\,+\,\kappa_{\lambda}''\; B(z)
\end{align}
where $\kappa_{\lambda}'' $ can be made smaller than $1$ by choosing $\lambda $ large, uniformly in $z=E+i\varepsilon$
and $B^{*}(z) $ is uniformly bounded in $\varepsilon $.
The assertion of the theorem now follows from \eqref{eq:estA} and \eqref{eq:estB2}.

\section{Appendix:(Positivity of Auxiliary Operators)}\label{sec:Appendix}

In this section we prove that $H_{0,\Gamma^{c}}=\chi_{\ZZ^{d}\setminus\Gamma}\,H_{0}\,\chi_{\ZZ^{d}\setminus\Gamma}$ as well as some other auxiliary operators are strictly positive. For our proof we use operators on $\ell^{2}(\ZZ^{d})$
with `Neumann boundary conditions', in the following sense:

\begin{defin}\label{def:Neumann}
   If $\Lambda$ is a (nonempty) subset of $\ZZ^{d} $ we define the Neumann Laplacian ${H_{0}}^{\Lambda}$ on $\ell^{2}(\Lambda)$ by
\begin{align}
 {H_{0}}^{\Lambda}\,u(n)~:=~\sum_{\myover{j\in\Lambda}{\| n-j \|=1}}\,\big(u(n)-u(j)\big)\,.
\end{align}
\end{defin}
For any $n\in\Lambda\subset\ZZ^{d}$ we define the \emph{coordination number} $\kappa_{\Lambda}(n)$ as the number of neighbors of $n$ within the set (graph) $\Lambda$, i.\,e.
\begin{align}
   \kappa_{\Lambda}(n)~:=~\#\{i\in\Lambda\mid |i-n|=1\}
\end{align}
and the \emph{boundary} $\partial^{-}\Lambda$ of $\Lambda$ by
\begin{align}
   \partial^{-}\Lambda~:=~\{i\in\Lambda\mid \exists_{j\not\in\Lambda}\, |i-j|=1\}\,.
\end{align}
With these notations we get
\begin{align}
   {H_{0}}^{\Lambda}\,u(n)~&=~\kappa_{\Lambda}(n)\,u(n)\,-\,\sum_{\myover{|j-n|=1}{j\in\Lambda}}u(j)\\
\intertext{while}\label{eq:simple}
    \chi_{\Lambda}{H_{0}}\chi_{\Lambda}\,u(n)~&=~2d\,u(n)\,-\,\sum_{\myover{|j-n|=1}{j\in\Lambda}}u(j)\\
\intertext{consequently}\label{eq:perturbation}
    \chi_{\Lambda}{H_{0}}\chi_{\Lambda}~&=~{H_{0}}^{\Lambda}\,+\,(2d-\kappa_{\Lambda})
\end{align}

We summarize some of the properties of the operators ${H_{0}}^{\Lambda}$ in the following Proposition.

\begin{prop}
   For any (nonempty) set $\Lambda\subset\ZZ^{d}$ we have
    \begin{enumerate}
        \item $\inf\sigma({H_{0}}^{\Lambda})~=~0$\,,
        \item If $\Lambda$ is finite, then $\inf\sigma({H_{0}}^{\Lambda})$ is an eigenvalue and $\varphi(n)=\frac{1}{|\Lambda|^{1/2}}$ is
        a normalized eigenfunction.
        \item If $\Lambda$ is finite and connected, then $\inf\sigma({H_{0}}^{\Lambda})$ is a simple eigenvalue.
        \item If $\Lambda$ is the disjoint union of the (nonempty) sets $\Lambda_{i} $ then
         \begin{align}\label{eq:super}
            {H_{0}}^{\Lambda}~\geq~\bigoplus {H_{0}}^{\Lambda_{i}}
            \end{align}
        \item $\chi_{\Lambda}{H_{0}}\chi_{\Lambda}~\geq~{H_{0}}^{\Lambda}$
    \end{enumerate}
\end{prop}

These properties of ${H_{0}}^{\Lambda} $, in particular the superadditivity \eqref{eq:super},  resemble characteristic properties of  Neumann Laplacians on $\RR^{d}$.
This is the reason we call ${H_{0}}^{\Lambda} $ the Neumann Laplacian, following Simon \cite{BS}.

It is tempting to think of $H_{0,A}$ as the Laplacian on $A$ with `Dirichlet' boundary conditions.
However, as was explained by Barry Simon in \cite{BS}, the operators $H_{0,A}$ are \emph{not} subadditive
as one should expect for an analog of Dirichlet conditions. Simon defines another type of boundary conditions which satisfy this property
   (see \cite{BS} or \cite{WK} for details). We speak of \emph{simple boundary conditions} when talking about operators as in \eqref{eq:simple}.

Below we will also combine Neumann and simple boundary conditions. So, if $\Lambda $ and $A$ are subsets of $\ZZ^{d} $ we consider the operator
\begin{align}
   \chi_{A}{H_{0}}^{\Lambda}\chi_{A}
\end{align}
as an operator on $\ell^{2}(A\cap\Lambda) $ and talk about Neumann conditions on $\Lambda $ and simple boundary conditions on $A$.

\textbf{Proof} (Proposition \ref{prop:positivity}){\bf:} Write $\ZZ^{d}$ as the disjoint union of cubes $\Lambda_{1}(i)$ with $i\in M\ZZ^{d}$
and set $A:=\Gamma^{c}$. Then
\begin{align}
   {H_{0,A}}~&=~\chi_{A}{H_{0}}\chi_{A}\\
&\geq~ \chi_{A}\;\bigoplus_{i\in M\ZZ_{d}} {H_{0}}^{\Lambda_{1}(i)}\;\chi_{A}\\
&=~\bigoplus_{i\in M\ZZ_{d}}~\Big(\chi_{A}\,{H_{0}}^{\Lambda_{1}(i)}\,\chi_{A}\Big)
\end{align}
Thus
\begin{align}
   \inf\sigma({H_{0,A}})~\geq~\inf\sigma\big(\chi_{A}\,{H_{0}}^{\Lambda_{1}}\,\chi_{A}\big)
\end{align}
Consequently, it suffices to prove that the operator $ \chi_{A}\,{H_{0}}^{\Lambda_{1}}\,\chi_{A}$ is strictly positive.

From \eqref{eq:perturbation} we have
\begin{align}
   \chi_{A}\,{H_{0}}^{\Lambda_{1}}\,\chi_{A}~=~{H_{0}}^{\Lambda_{1}\cap A}\,+\,(2d-\kappa_{\Lambda_{1}\cap A})
\end{align}

By Assumption \ref{assumG} $\Lambda_{1}\setminus A\neq\emptyset$, hence $q(n):=2d-\kappa_{\Lambda_{1}\cap A}(n)\geq 1$
for some $n\in\Lambda_{1}\cap A$.

Setting $B(t)={H_{0}}^{\Lambda_{1}\cap A}+\,t\,q$ and $e(t)=\inf\sigma(B(t))$ we have $e(0)=0$ and by
the Hellmann-Feynman-Theorem
\begin{align}
   e'(0)~=~\sum_{n\in\Lambda_{1}\cap A}\,q(n)~>~0\,,
\end{align}
thus
\begin{align}
   e(1)~=~\inf\sigma\big(\chi_{A}\,{H_{0}}^{\Lambda_{1}}\,\chi_{A}\big)~>~0
\end{align}
\qed

The above proof does not use Assumption \ref{assumG} in its full strength. All what is needed is that $\Lambda_{1}\cap \Gamma \neq \emptyset$, which
is true for any \emph{Delone set}.

\end{document}